\documentclass[journal,twoside,web]{ieeecolor}
\usepackage{generic}
\usepackage{cite}
\usepackage{amsmath,amssymb,amsfonts}
\usepackage{algorithmic}
\usepackage{graphicx}
\usepackage{algorithm,algorithmic}
\usepackage{hyperref}
\hypersetup{hidelinks=true}
\usepackage{textcomp}
\usepackage{booktabs}	% manual add
\usepackage{multirow}	% manual add
\usepackage{tabularx}   % manual add
\usepackage{makecell}   % manual add
\usepackage[caption=false,font=footnotesize]{subfig}		% manual add
\def\BibTeX{{\rm B\kern-.05em{\sc i\kern-.025em b}\kern-.08em
    T\kern-.1667em\lower.7ex\hbox{E}\kern-.125emX}}
\begin{document}
\title{SomnoNet: A Lightweight and Interpretable Framework for Sleep Staging Using Single-Channel EEG}
\author{Shengwei~Guo and Guobing~Sun
\thanks{Shengwei~Guo and Guobing~Sun are with the Key Laboratory of Information Fusion Estimation and Detection, School of Electronic Engineering, Heilongjiang University, Harbin 150080, China.}
\thanks{Shengwei~Guo (e-mail: 2211849@s.hlju.edu.cn).}
\thanks{Corresponding author: Guobing~Sun (e-mail: sunguobing@hlju.edu.cn).}
}

\maketitle

\begin{abstract}
	Automated sleep staging from single-channel electroencephalography (EEG) is attractive for scalable sleep assessment, but practical systems must jointly address accuracy, efficiency, and clinical interpretability. We propose SomnoNet, a hierarchical raw-EEG framework motivated by expert scoring practice. The model first extracts multi-scale local rhythm representations from short temporal chunks and then integrates intra-epoch organization and inter-epoch context using hierarchical temporal modeling. On two large public benchmarks, SomnoNet achieves 80.9\% accuracy, 79.0\% macro-F1, and 0.739 kappa on Physio2018, and 88.0\% accuracy, 80.7\% macro-F1, and 0.831 kappa on SHHS. To support resource-constrained deployment, we further develop SomnoNet-Nano, a frozen-encoder compact variant that reuses the learned morphology encoder and replaces the original temporal stack with a lightweight sequence unit. SomnoNet-Nano contains 0.049M parameters, runs in 29.49 ms per 30-s epoch on an i7-12700F CPU under FP32 inference, and retains 99.5\% and 99.3\% of the full-model accuracy on Physio2018 and SHHS, respectively. Finally, rhythm-aware decision analysis visualizes segment-level model evidence and relates predictions to clinically meaningful EEG patterns. These results suggest that SomnoNet balances predictive performance, compactness, and transparent decision support for single-channel EEG sleep staging.
\end{abstract}

\begin{IEEEkeywords}
Sleep Staging, Single-Channel EEG, Lightweight Neural Networks, Model Interpretability, Digital Health, Artificial Intelligence
\end{IEEEkeywords}

	\section{Introduction}\label{sec1}

Sleep is a fundamental physiological process that underpins human health, cognitive performance, and emotional regulation\cite{siegel2005clues, roleSleep}. Accurate identification of sleep stages through polysomnography (PSG), especially electroencephalography (EEG), is central to both scientific research and clinical practice\cite{mahowald2005, AASM}. Reliable staging is essential for diagnosing and managing common disorders such as insomnia, sleep apnea, and narcolepsy. However, manual scoring remains labor-intensive, time-consuming, and prone to inter-scorer variability, making it unsuitable for large-scale or long-term monitoring. This creates an urgent need for automatic methods that are both reliable and scalable.

\begin{figure}[!t]
	\centering
	\subfloat[]{\includegraphics[width=3.3in]{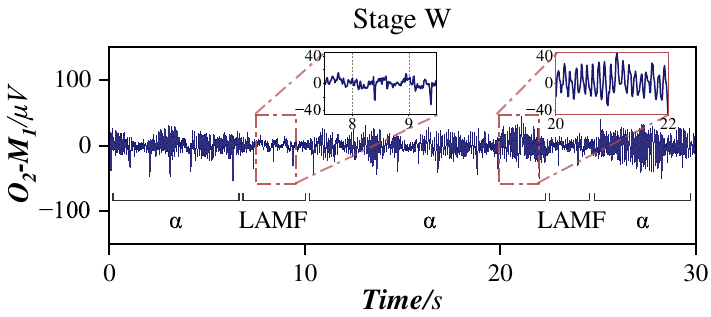}
		\label{fig:oriEEG_W}}
	\hfil
	\subfloat[]{\includegraphics[width=3.3in]{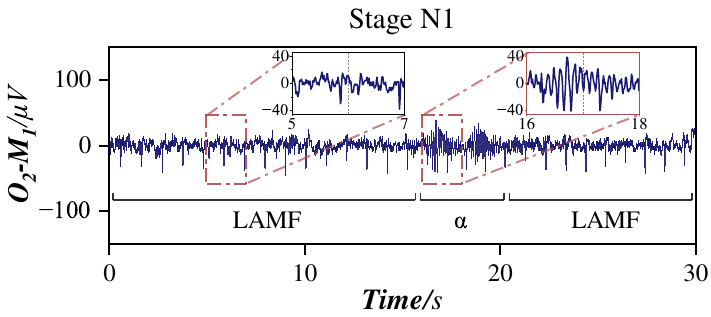}
		\label{fig:oriEEG_N1}}
	\caption{Clinical illustration of AASM alpha-rhythm criteria using representative EEG signals from the $O_2$--$M_1$ channel of the Physio2018 dataset: (a) wake (W), (b) Stage N1. The occipital lead is used here only for visualization of posterior $\alpha$ activity; the experimental channels are reported in Table~\ref{tab:dataset_1}. Each panel corresponds to a 30-second epoch. The red dotted boxes mark magnified representative signal segments. In (a), dominant $\alpha$ activity ($>$50\% of the epoch) is consistent with wakefulness. In (b), $\alpha$ activity decreases below 50\%, giving way to low-amplitude mixed-frequency (LAMF) activity, which is consistent with Stage N1.}
	\label{fig:oriEEG}
\end{figure}

In recent years, deep learning has demonstrated remarkable potential for automated sleep staging. Unlike traditional machine learning approaches, neural network-based models can directly learn discriminative representations from raw or minimally preprocessed EEG signals, capturing nonlinear temporal patterns and achieving strong accuracy and robustness. Representative frameworks include DeepSleepNet\cite{DeepSleepNet}, TinySleepNet\cite{TinySleepNet}, IITNet\cite{IITNet}, U-Time\cite{U-time}, SeqSleepNet\cite{SeqSleepNet}, XSleepNet\cite{XSleepNet}, SleepTransformer\cite{Sleeptransformer}, and SleePyCo\cite{Sleepyco}. More recent studies such as U-Sleep\cite{U-Sleep}, DistillSleep\cite{park2025distillsleep}, and SLEEPYLAND\cite{rossi2025sleepyland} further emphasize the importance of efficiency, interpretability, and fair evaluation. Collectively, these studies demonstrate that both local waveform morphology and temporal context are important for sleep staging.

Despite this progress, three challenges still limit the translation of automatic sleep staging models into practical biomedical and health informatics systems. First, many high-performing models are evaluated mainly by aggregate accuracy, whereas their evidence at the level of clinically meaningful EEG rhythms remains insufficiently characterized. Second, lightweight models are often obtained by simply shrinking the architecture or by using teacher--student distillation, leaving unclear which component carries transferable sleep-stage information. Third, comparisons across studies are complicated by heterogeneous channel selections, preprocessing pipelines, and evaluation protocols. These challenges motivate a framework that is not only accurate, but also structurally interpretable, compact, and explicit about its evaluation assumptions.

In this work, we propose \textbf{SomnoNet}, a hierarchical end-to-end framework for single-channel EEG sleep staging. The model is motivated by how sleep experts inspect EEG: characteristic graphoelements are first recognized within short temporal segments, and the final stage decision is then made by considering both intra-epoch organization and inter-epoch context. Following this rationale, SomnoNet combines a multi-scale convolutional encoder for local rhythm representation with hierarchical temporal modeling for intra-epoch and inter-epoch sequence learning. This design enables the model to capture both local EEG morphology and longer-range transition context in a unified framework.

To support resource-constrained deployment, we further introduce \textbf{SomnoNet-Nano}, a compact variant that reuses the morphology encoder learned by the full model and replaces the original hierarchical temporal stack with a lightweight sequence unit. The full SomnoNet serves as a performance-oriented reference architecture and interpretability carrier, whereas SomnoNet-Nano serves as the deployment-oriented compact variant. In this sense, SomnoNet-Nano is not only a compressed model, but also an empirical test of structural efficiency: if a frozen encoder can support a much smaller temporal head with minimal performance loss, then the proposed encoder has learned transferable local EEG morphology rather than merely overfitting the full temporal stack.

In addition, we incorporate a rhythm-aware interpretation strategy to relate model decisions to characteristic EEG patterns. Instead of treating interpretability as a separate prediction system, we use it as an analysis layer that presents model evidence at different levels of abstraction, thereby improving clinical transparency and facilitating expert inspection.

The main contributions of this work can be summarized as follows:

\begin{enumerate}
	\item \textbf{Clinically motivated hierarchical raw-EEG modeling.} We propose SomnoNet, a single-channel EEG framework that decomposes sleep staging into local rhythm representation, intra-epoch organization modeling, and inter-epoch contextual aggregation, reflecting the multi-scale reasoning process of manual sleep scoring.
	\item \textbf{Frozen-encoder compact transfer.} We develop SomnoNet-Nano, a compact variant that reuses the learned morphology encoder and replaces the original temporal stack with a lightweight sequence unit. The frozen-encoder results provide empirical evidence that local EEG morphology can be represented compactly and transferred to a much smaller temporal head.
	\item \textbf{Rhythm-aware decision analysis.} We introduce a multi-level attribution strategy that visualizes segment-level model evidence and relates model decisions to clinically meaningful EEG rhythms and temporal contexts.
\end{enumerate}

Overall, this work aims to bridge predictive performance, computational efficiency, and clinical interpretability in automatic single-channel EEG sleep staging.

\begin{figure}[!t]
	\centering
	\includegraphics[width=.85\linewidth]{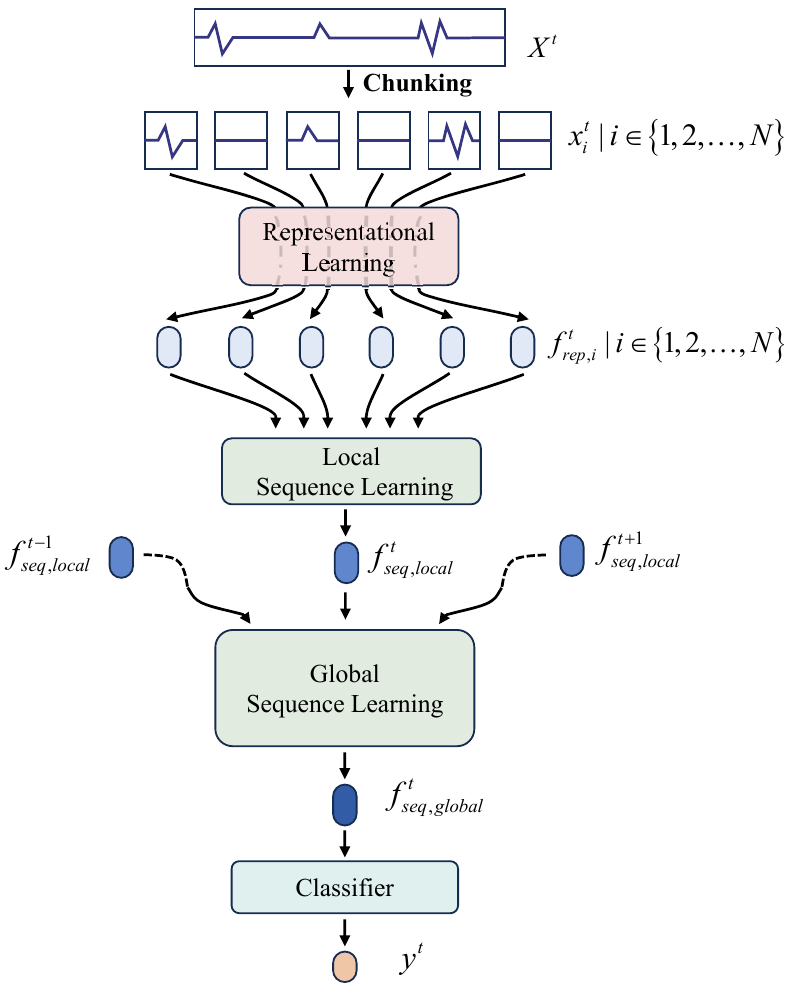}
	\caption{Overall architecture of SomnoNet. Each 30-second EEG epoch is divided into $N$ sub-segments and encoded by the representation learning module to capture characteristic rhythms. A hierarchical sequence learning module then integrates local (intra-epoch) and global (inter-epoch) modeling to simultaneously capture short-term dynamics and longer-range sleep transitions. Finally, a classifier outputs the stage label for the current epoch.}
	\label{fig:over_arch}
\end{figure}

\section{Method}

\subsection{SomnoNet}

\subsubsection{Overall Architecture}

Sleep experts do not determine sleep stages from isolated signal samples. Instead, they identify characteristic graphoelements and rhythm patterns within short temporal segments of a 30-second epoch, and then combine these local observations with broader contextual information to make the final stage decision. Fig.~\ref{fig:oriEEG} provides a clinical illustration using two raw EEG segments from the $O_2$--$M_1$ channel of the Physio2018 dataset\cite{goldberger2000physiobank, physio2018}: Fig.~\ref{fig:oriEEG_W} corresponds to wakefulness (Stage W), whereas Fig.~\ref{fig:oriEEG_N1} corresponds to Stage N1. This occipital channel is shown only to illustrate the AASM alpha-rhythm criterion, because posterior alpha activity is typically more visually prominent; the experimental single-channel inputs are the central leads specified in Section~\ref{sec:materials}. According to the AASM manual, Stage W is characterized by pronounced alpha activity (8--13 Hz) in the occipital region. When the proportion of $\alpha$ rhythm falls below 50\% and is replaced by low-amplitude mixed-frequency (LAMF) activity, the subject transitions into Stage N1. In practice, human experts rely on both rhythm morphology and temporal duration: in Fig.~\ref{fig:oriEEG_W}, dominant $\alpha$ activity indicates Stage W, whereas in Fig.~\ref{fig:oriEEG_N1}, reduced $\alpha$ activity with prevailing LAMF is consistent with Stage N1.

Motivated by this scoring process, we propose SomnoNet, an end-to-end framework for single-channel EEG-based sleep staging. As illustrated in Fig.~\ref{fig:over_arch}, SomnoNet directly maps raw EEG to sleep stage categories through a hierarchical architecture comprising five modules:
\begin{itemize}
	\item \textbf{Chunking:} each EEG epoch $x^t$ is divided into $N$ consecutive temporal chunks $\{x_{i}^{t} \mid i=1,\ldots,N\}$ to retain fine-grained local dynamics.
	\item \textbf{Representation learning:} a convolutional network extracts chunk-level features $f_{rep,i}^{t}$, capturing local rhythm patterns.
	\item \textbf{Intra-epoch sequence learning:} a recurrent layer models dependencies among chunk features, producing an epoch-level representation $f_{seq,local}^{t}$ that encodes short-term temporal context.
	\item \textbf{Inter-epoch sequence learning:} higher-level recurrent layers capture long-range dependencies across consecutive epochs, yielding global temporal features $f_{seq,global}^{t}$.
	\item \textbf{Classifier:} a fully connected layer maps global features to the predicted stage label $y^t$, producing the final sleep staging decision.
\end{itemize}

As illustrated in Fig.~\ref{fig:over_arch}, this hierarchical design progressively extracts local features, models both intra-epoch and inter-epoch dependencies, and captures global temporal dynamics. By combining raw-EEG end-to-end learning with clinically motivated temporal modeling, SomnoNet provides a compact and interpretable architecture for single-channel sleep staging.

\subsubsection{Representation Learning}

The representation module comprises three Multi-scale Convolutional Feature Extraction Modules (MCFEMs) followed by a global average pooling layer, designed to efficiently capture discriminative EEG features across temporal scales \cite{he2016res, dosovitskiy2020Vit, long2015fully}. Since sleep stages are characterized by distinct rhythmic patterns, multi-scale modeling is essential for robust staging.

\begin{figure}[!t]
	\centering
	\includegraphics[width=.9\linewidth]{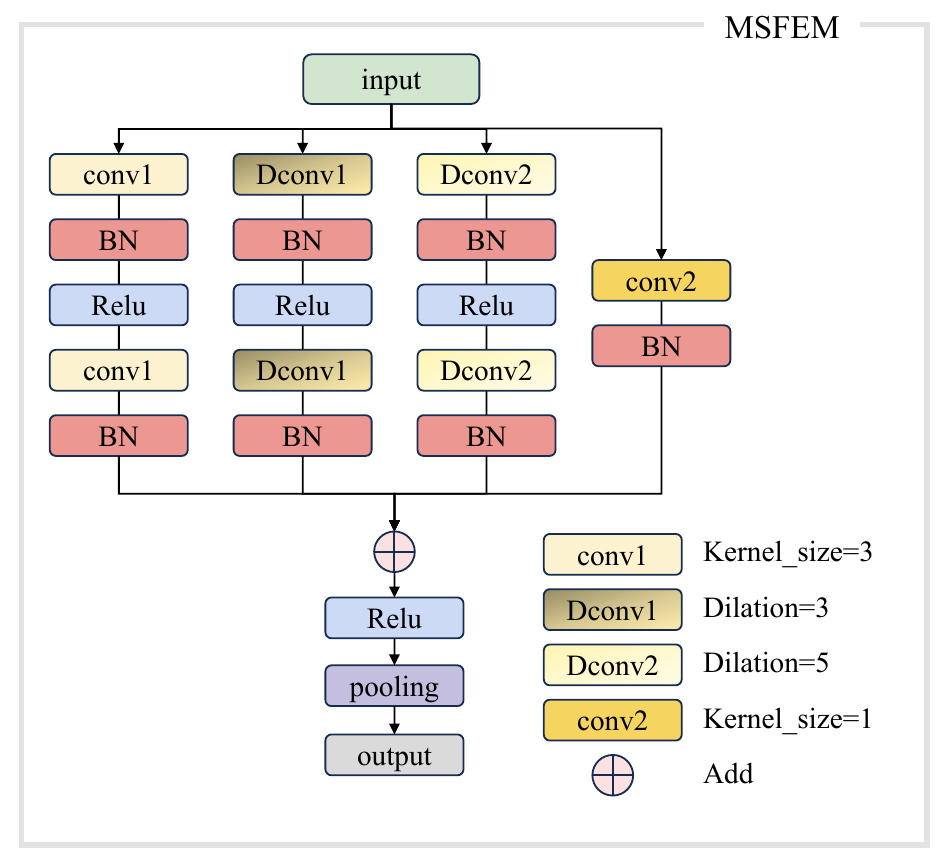}
	\caption{Multi-scale Convolutional Feature Extraction Module (MCFEM). The input EEG segment is processed through multiple parallel convolutions with different receptive fields to capture diverse temporal patterns. The extracted feature maps are fused via residual connections, followed by activation and pooling, enabling effective learning and integration of features across multiple temporal scales.}
	\label{fig:MSFEM}
\end{figure}

As illustrated in Fig.~\ref{fig:MSFEM}, each MCFEM applies a $3\times1$ convolution to extract short-term dynamics. Dilated convolutions (dilation rates 3 and 5) expand the receptive field without a substantial increase in parameter count, capturing broader temporal dependencies. A subsequent $1\times1$ convolution integrates features across channels, enabling interaction among multi-scale representations. Each convolution is followed by batch normalization and nonlinear activation for training stability, and pooling layers compress features while preserving discriminative information.

The multi-scale design is particularly important for sleep EEG because discriminative patterns may appear at different temporal extents and frequency ranges. By combining standard and dilated convolutions, the encoder captures both transient graphoelements and broader rhythm structures at low computational cost. This also improves robustness across datasets with different sampling rates, since characteristic EEG rhythms can still be represented through receptive fields spanning multiple temporal scales.

\subsubsection{Sequence Learning and Classification}

Sleep staging depends not only on local waveform morphology but also on the temporal organization of events within an epoch and the contextual continuity across neighboring epochs. To capture these two levels of temporal dependency, SomnoNet adopts a hierarchical sequence learning design.
\begin{itemize}
	\item \textbf{Intra-epoch level:} a bidirectional gated recurrent unit (Bi-GRU) models dependencies among chunk-level features within the same epoch, producing $f_{seq,local}^{t}$. This module captures the ordering, duration, and distribution of transient graphoelements and rhythmic segments within the epoch, which is clinically important because the temporal location of characteristic patterns may affect stage interpretation.
	\item \textbf{Inter-epoch level:} a stack of Bi-GRU layers captures temporal dependencies across consecutive epochs, yielding $f_{seq,global}^{t}$. This module models evolving transitions between sleep stages and provides broader context for ambiguous cases.
\end{itemize}

We adopt Bi-GRU as the temporal modeling unit not because it is universally superior to newer sequence models, but because it provided the best empirical accuracy among the tested alternatives in our single-channel EEG setting under a compact parameter budget. The bidirectional design is also consistent with standard offline sleep staging, where neighboring epochs are jointly considered during scoring.

Finally, a lightweight multilayer perceptron (MLP) classifier maps the global representation to the final stage label $y^t$. This single fully connected layer preserves computational efficiency while maintaining end-to-end discriminative capability.

To formalize the end-to-end inference, consider an input of consecutive EEG epochs, each segmented into $N$ chunks. The overall process can be expressed as Eq.~(\ref{eq:net}), where $\operatorname{Chunk}(\cdot)$ denotes segmentation, $\operatorname{Rep}(\cdot)$ the MCFEM-based representation module, $\operatorname{LocalSeq}(\cdot)$ and $\operatorname{GlobalSeq}(\cdot)$ the intra- and inter-epoch sequence models, and $\operatorname{Classifier}(\cdot)$ the final prediction layer.

\begin{equation}
		\begin{aligned}
			& \{x_i^t\}_{i=1}^N = \operatorname{Chunk}(x^t) \\
			& f_{rep,i}^t = \operatorname{Rep}(x_i^t),\quad i=1,\dots,N\\
			& f_{seq,local}^t = \operatorname{LocalSeq}\left(\{f_{rep,i}^t\}_{i=1}^N\right)\\
			& f_{seq,global}^t = \operatorname{GlobalSeq}\big( \ldots, f_{seq,local}^{t-1}, f_{seq,local}^{t}, f_{seq,local}^{t+1}, \ldots \big)\\
			& y^t = \operatorname{Classifier}(f_{seq,global}^t)
		\end{aligned}
		\label{eq:net}
\end{equation}

This hierarchical formulation enables SomnoNet to capture EEG dynamics across multiple temporal scales, effectively integrating local fine-grained details with global context to achieve accurate and efficient sleep staging.

\subsection{SomnoNet-Nano}

Our preliminary analysis showed that most parameters of SomnoNet are concentrated in the temporal modeling modules, whereas the convolutional encoder remains relatively compact. This observation motivated a targeted question: can the encoder learn local EEG morphology that remains useful after the temporal head is aggressively compressed?

\begin{figure}[!t]
	\centering
	\includegraphics[width=.85\linewidth]{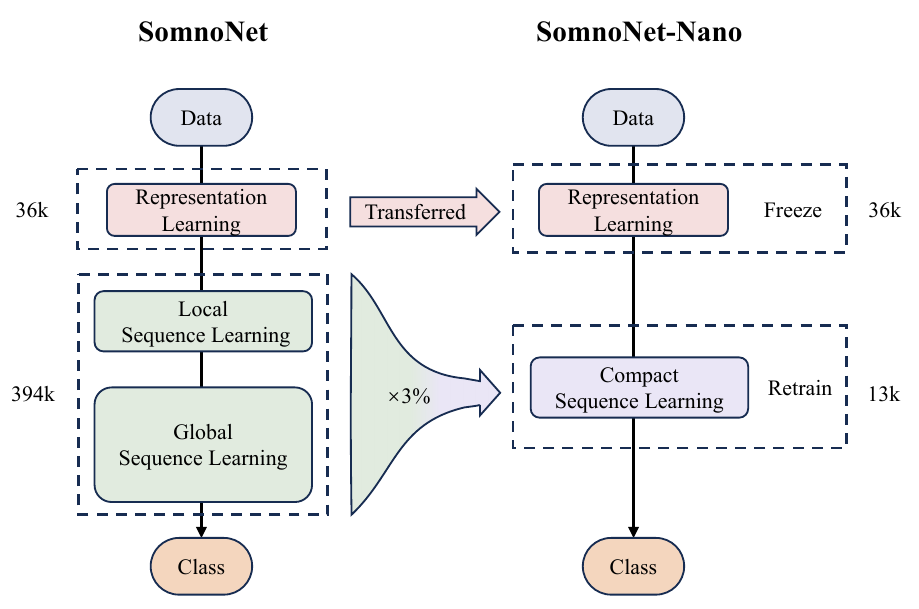}
	\caption{Training scheme of SomnoNet-Nano. SomnoNet-Nano reuses the representation learning module from SomnoNet with frozen weights, while replacing the original local and global sequence learning modules with a lightweight mixed sequence learning module. Consequently, only the compact temporal head is trained, greatly reducing optimization cost and model size.}
	\label{fig:net_nano}
\end{figure}

Based on this analysis, we propose SomnoNet-Nano, a deployment-oriented compact variant of SomnoNet that freezes the representation learning module and trains only a lightweight mixed sequence learning module, as illustrated in Fig.~\ref{fig:net_nano}. This design keeps the manuscript narrative separated: the full SomnoNet is the performance-oriented reference model, whereas SomnoNet-Nano is the practical compact model for resource-constrained deployment.

Its core design includes:
\begin{itemize}
	\item \textbf{Frozen encoder:} the representation module learned by the full model is retained with frozen weights, allowing the compact model to reuse robust local morphological features.
	\item \textbf{Compact Sequence Unit (CSU):} the original local and global sequence modules are replaced by a unified single-layer Bi-GRU, preserving essential temporal modeling capacity while simplifying the architecture.
	\item \textbf{Compact parameter footprint:} the original SomnoNet contains approximately 430k parameters, with about 394k in sequence modeling. In comparison, the CSU in SomnoNet-Nano requires only about 13k trainable parameters, reducing the total model size to about 49k parameters while preserving strong predictive performance.
\end{itemize}

\begin{equation}
		\begin{aligned}
			& \{x_i^t\}_{i=1}^N = \operatorname{Chunk}(x^t) \\
			& f_{rep,i}^t = \operatorname{Rep}(x_i^t),\quad i=1,\dots,N\\
			& f_{seq,nano}^{t}=\operatorname{CompactSeq}\big(\{f_{rep,i}^{\tau }\}_{\tau \in T,\,i=1,\ldots ,N}\big)\\
			& y^t = \operatorname{Classifier}(f_{seq,nano}^t) 
		\end{aligned}
		\label{eq:nano}
\end{equation}

Formally, the end-to-end computation can be represented as in Eq.~(\ref{eq:nano}), where $\operatorname{CompactSeq}(\cdot)$ denotes the streamlined sequence modeling unit and $T$ denotes the temporal context window used for the target epoch.

This design is not merely a compression trick. Rather, it also provides evidence that the local representations learned by the full model are highly transferable. In our additional experiments, the frozen-encoder Nano model retained nearly all of the full-model accuracy, whereas the jointly trained Nano model showed a larger performance drop. This suggests that, under a very small parameter budget, decoupling local morphology learning from compact temporal modeling may be more effective than jointly optimizing both from scratch.

\section{Materials}\label{sec:materials}

We conducted experiments on two widely used benchmark datasets for sleep staging: Physio2018\cite{goldberger2000physiobank, physio2018} and SHHS\cite{SHHS, SHHS_2}. To make the evaluation comparable with XSleepNet and subsequent single-channel EEG baselines, we followed the same dataset preparation and subject-level evaluation protocol where applicable. Importantly, this alignment concerns the data protocol rather than the model pipeline: unlike XSleepNet, which jointly learns from raw signals and time--frequency images, SomnoNet uses only raw single-channel EEG as input.

Each PSG recording was segmented into non-overlapping 30-second epochs according to the provided hypnogram annotations. Only epochs labeled as W, N1, N2, N3, and REM were retained. For SHHS, R\&K Stages 3 and 4 were merged into N3 to align with the AASM five-stage taxonomy, and Movement/Unknown epochs were excluded. No handcrafted spectral features or spectrogram inputs were used by SomnoNet. All partitions were performed at the subject level to avoid subject leakage.

We selected central EEG leads because they are positioned between the frontal and occipital regions and can capture representative rhythms from multiple cortical areas while remaining practical for single-channel acquisition. The Physio2018 experiments used the $C_3$--$A_2$ channel, and the SHHS experiments used the $C_4$--$M_1$ channel. The $O_2$--$M_1$ signals shown in Fig.~\ref{fig:oriEEG} are used only as a clinical illustration of posterior alpha activity and are not the experimental input channel.

Table~\ref{tab:dataset_1} summarizes the key EEG characteristics of the datasets, and Table~\ref{tab:dataset_2} reports the distribution of samples across sleep stages. All subsequent experiments and evaluations were conducted on the preprocessed data described above.

\begin{table}[!t]
	\centering
	\caption{Dataset distribution information for Physio2018 and SHHS.}
	\begin{tabular}{cccc}
		\toprule
		Dataset    & Scoring manual & Subjects & Channel \\
		\midrule
		Physio2018 & AASM           & 994    & ${{\text{C}}_{\text{3}}}\text{-}{{\text{A}}_{\text{2}}}$   \\
		SHHS       & R\&K           & 5793   & ${{\text{C}}_{\text{4}}}\text{-}{{\text{M}}_{\text{1}}}$   \\
		\bottomrule
	\end{tabular}
	\label{tab:dataset_1}
\end{table}

\begin{table}[!t]
	\centering
	\caption{Number of samples per sleep stage in the datasets.}
	\setlength{\tabcolsep}{3pt} 
	\begin{tabular}{ccccccc}
		\toprule
		\multirow{2}{*}{Dataset} & \multicolumn{6}{c}{Sleep epoch category distribution} \\
		\cmidrule(lr){2-7}
		& W & N1 & N2 & N3 & R & Total \\
		\midrule
		\multirow{2}{*}{Physio2018} 
		& 157,945 & 136,978 & 377,870 & 102,592 & 116,877 & 892,262 \\
		& (17.7\%) & (15.4\%) & (42.3\%) & (11.5\%) & (13.1\%) & (100\%) \\
		\midrule
		\multirow{2}{*}{SHHS} 
		& 1,691,288 & 217,583 & 2,397,460 & 739,403 & 817,473 & 5,863,207 \\
		& (28.8\%) & (3.7\%) & (40.9\%) & (12.6\%) & (13.9\%) & (100\%) \\
		\bottomrule
	\end{tabular}
	\label{tab:dataset_2}
\end{table}

\section{Experiments}

\subsection{Configuration and Preparation}

All experiments were conducted on the preprocessed datasets described in Section~\ref{sec:materials}.

For the evaluation protocol, the experimental settings were aligned with prior work for both datasets. For Physio2018, we followed the established five-fold subject-independent cross-validation scheme used in XSleepNet-style benchmarking. For SHHS, the dataset was split at the subject level into training and testing sets using a 7:3 ratio, consistent with the protocol reported in \cite{XSleepNet}. In all experiments, validation data were drawn only from the training portion of each split, ensuring that the test subjects were never used for model selection.

Model training was conducted on an NVIDIA RTX 4090 GPU (24 GB). Unless otherwise specified, the batch size was set to 32, with a maximum of 150 epochs. Early stopping was applied when the validation loss failed to decrease for 8 consecutive epochs, preventing overfitting and reducing unnecessary computation. Unless otherwise noted, the reported training and evaluation results were obtained on this platform. The latency benchmarks reported later in Table~\ref{tab:efficiency} were measured separately with batch size 1 under FP32 precision on both an NVIDIA RTX 4090 GPU and an Intel Core i7-12700F CPU. No mixed-precision acceleration was used for either GPU or CPU timing.

For optimization, we employed AdamW\cite{FixingAdam} with weight decay $1\times10^{-2}$. The initial learning rate was set to $1\times10^{-4}$ and scheduled using OneCycleLR with \texttt{max\_lr} $=3\times10^{-4}$, \texttt{pct\_start} $=0.1$, \texttt{div\_factor} $=10.0$, and \texttt{final\_div\_factor} $=100.0$. Cross-entropy loss was used to quantify the discrepancy between predicted and ground-truth distributions.

Unless otherwise specified, mechanistic analyses including sampling-rate robustness, computational efficiency, and interpretability evaluation were conducted on the first Physio2018 fold, while the main performance comparison reports the cross-validation results.

\subsection{Results and Analysis}

\subsubsection{Performance of SomnoNet}

\begin{table*}[!t]
	\centering
	\caption{Literature-level comparison with representative single-channel EEG sleep staging methods. Because preprocessing, channel selection, and data splitting may differ across studies, these results are used to contextualize performance rather than to claim a strictly controlled head-to-head benchmark. OA, MF1, and $\kappa$ denote overall accuracy, macro-F1, and Cohen's kappa, respectively. Boldface highlights the best reported value in each column, and underlined values indicate the second-best reported value.}
	\setlength{\tabcolsep}{3pt} 
	\begin{tabular}{cccccccccc}
		\toprule
		\multirow{2}{*}{Dataset} & \multirow{2}{*}{Method} 
		& \multicolumn{3}{c}{Overall} 
		& \multicolumn{5}{c}{F1 score} \\
		\cmidrule(lr){3-5} \cmidrule(lr){6-10}
		& & OA & MF1 & $\kappa$ & W & N1 & N2 & N3 & R \\
		\midrule
		\multirow{6}{*}{Physio2018} 
		& SomnoNet (ours)        & \textbf{80.9} & \textbf{79.0} & \textbf{0.739} & \textbf{84.6} & \underline{59.0} & \underline{85.1} & \textbf{80.2} & \textbf{86.3} \\
		& SleePyCo\cite{Sleepyco}        & \textbf{80.9} & \underline{78.9} & \underline{0.737} & \underline{84.2} & \textbf{59.3} & \textbf{85.3} & \underline{79.4} & \textbf{86.3} \\
		& Distill-T\cite{park2025distillsleep} & \underline{80.4} & \underline{78.9} & 0.733 & -- & -- & -- & -- & -- \\
		& XSleepNet\cite{XSleepNet}       & 80.3 & 78.6 & 0.732 & -- & -- & -- & -- & -- \\
		& SeqSleepNet\cite{SeqSleepNet}     & 79.4 & 77.6 & 0.719 & -- & -- & -- & -- & -- \\
		& U-Time\cite{U-time}          & 78.8 & 77.4 & 0.714 & 82.5 & \underline{59.0} & 83.1 & 79.0 & \underline{83.5} \\
		\midrule
		\multirow{7}{*}{SHHS} 
		& SomnoNet (ours)        & \textbf{88.0} & \underline{80.7} & \textbf{0.831} & \textbf{92.9} & 48.5 & \textbf{88.5} & 84.8 & \textbf{88.7} \\
		& SleePyCo\cite{Sleepyco}        & \underline{87.9} & \underline{80.7} & \underline{0.830} & \underline{92.6} & \underline{49.2} & \textbf{88.5} & 84.5 & \underline{88.6} \\
		& SleepTransformer\cite{Sleeptransformer} & 87.7 & 80.1 & 0.828 & 92.2 & 46.1 & 88.3 & \textbf{85.2} & \underline{88.6} \\
		& XSleepNet\cite{XSleepNet}       & 87.6 & \underline{80.7} & 0.826 & 92.0 & \textbf{49.9} & 88.3 & \underline{85.0} & 88.2 \\
		& Distill-T\cite{park2025distillsleep} & 86.8 & \textbf{81.1} & 0.814 & -- & -- & -- & -- & -- \\
		& IITNet\cite{IITNet}          & 86.7 & 79.8 & 0.812 & 90.1 & 48.1 & \underline{88.4} & \textbf{85.2} & 87.2 \\
		& SeqSleepNet\cite{SeqSleepNet}     & 86.5 & 78.5 & 0.810 & -- & -- & -- & -- & -- \\
		\bottomrule
	\end{tabular}
	\label{tab:net_result}
\end{table*}

Table~\ref{tab:net_result} summarizes the performance of SomnoNet compared with recent representative models. SomnoNet achieves the best or tied-best OA and $\kappa$ on both datasets while directly processing raw EEG through a compact end-to-end architecture. On SHHS, Distill-T reports a slightly higher MF1, indicating that SomnoNet should be interpreted as offering a favorable balance among accuracy, efficiency, and interpretability rather than uniformly dominating every metric. Compared with multi-view or distillation-heavy pipelines, SomnoNet keeps the input and training pipeline comparatively simple while maintaining competitive performance.

\begin{figure}[!t]
	\centering
	\includegraphics[width=.95\linewidth]{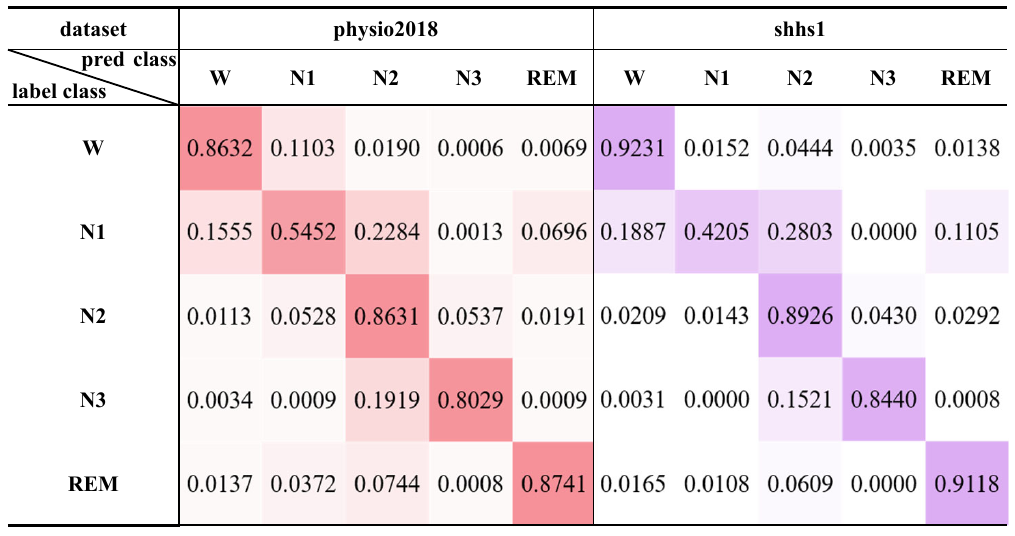}
	\caption{Confusion matrices of SomnoNet on the Physio2018 and SHHS datasets.}
	\label{fig:matrix}
\end{figure}

Fig.~\ref{fig:matrix} shows the confusion matrices of SomnoNet on both datasets, illustrating its classification behavior across sleep stages. These matrices clearly depict true positives, false positives, and error patterns, providing insight into per-stage performance.

\begin{figure}[!t]
	\centering
	\includegraphics[width=.65\linewidth]{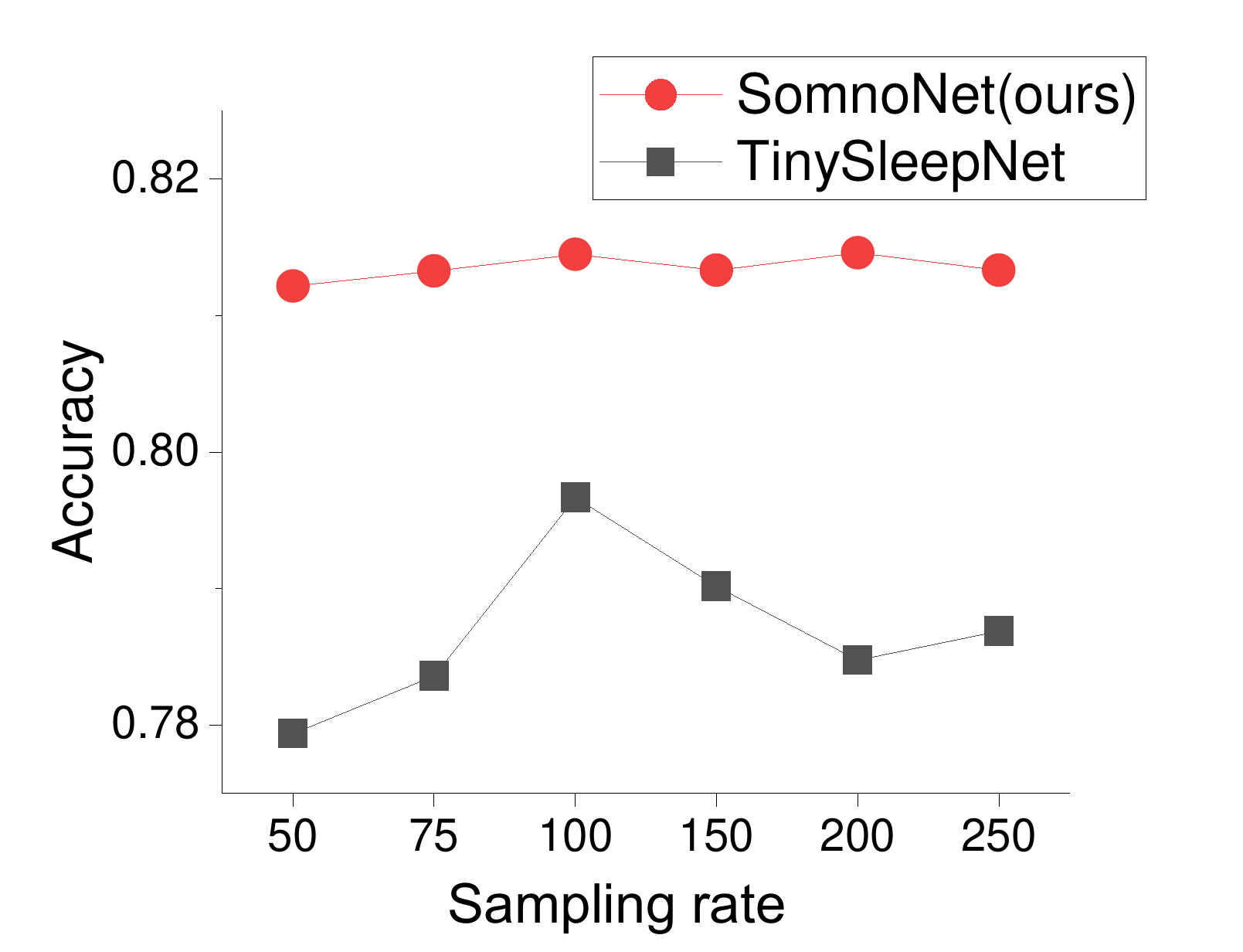}
	\caption{Comparison of model performance under different EEG sampling-rate settings: SomnoNet (ours) vs. TinySleepNet.}
	\label{fig:diff_sam}
\end{figure}

To assess generalization across different EEG sampling rates and frequency ranges, we conducted experiments presented in Fig.~\ref{fig:diff_sam}. TinySleepNet was used as a baseline, as its original implementation requires modifying convolution kernel sizes whenever the sampling rate changes, thereby changing the architecture and number of parameters.

In contrast, SomnoNet's multi-scale feature extraction module enables the architecture to remain unchanged across sampling rates, without manual adjustment. SomnoNet also maintains consistently strong performance across diverse conditions, validating its robustness and practical applicability.

Despite the overall strong results, the classification of Stage N1 remains challenging, which is consistent with observations across prior literature. Several factors contribute to this difficulty:

\textbf{Annotation bias:}
AASM scoring hinges on the proportion of occipital $\alpha$ rhythm. Small deviations around the 50\% threshold can lead to labeling inconsistencies.

\textbf{Absence of $\alpha$ rhythm:}
Roughly 10\% of individuals exhibit little or no $\alpha$ activity, requiring additional cues such as EMG, whose interpretation may vary between scorers.

\textbf{Central EEG leads:}
This study uses central channels ($C_3$--$A_2$ for Physio2018 and $C_4$--$M_1$ for SHHS). Although central leads capture broad cortical activity and are widely used in benchmark settings, posterior $\alpha$ rhythm is typically more prominent in occipital leads, which may make W--N1 transitions less visually separable in single-channel central EEG.

\subsubsection{Performance of SomnoNet-Nano}

Table~\ref{tab:nano_acc} compares the full model, the frozen-encoder Nano variant, and a jointly trained Nano variant. The frozen-encoder Nano model retains nearly all of the full-model accuracy, while the jointly trained compact model shows a larger drop. This pattern supports the efficiency of the proposed architecture, especially the encoder: once robust local morphological representations are learned, they can be reused effectively by a highly compact temporal head. By contrast, directly training a much smaller temporal model jointly appears less effective, likely because compact temporal modeling and local representation learning become harder to balance under a very small parameter budget.

\begin{table}[!t]
	\centering
	\caption{Accuracy comparison among SomnoNet and compact variants. ``SomnoNet-Nano'' denotes frozen-encoder training, and ``SomnoNet-Nano-JT'' denotes joint training from scratch. Retained performance is measured relative to the full model on the same dataset.}
	\setlength{\tabcolsep}{3pt} 
	\begin{tabular}{cccc}
		\toprule
		Dataset & Method & Accuracy & Retained performance \\
		\midrule
		\multirow{3}{*}{Physio2018} & SomnoNet-Nano & 80.5 & 99.5\% \\
		& SomnoNet-Nano-JT & 79.7 & 98.5\% \\
		& SomnoNet & 80.9 & 100\% \\
		\midrule
		\multirow{3}{*}{SHHS} & SomnoNet-Nano & 87.4 & 99.3\% \\
		& SomnoNet-Nano-JT & 87.2 & 99.1\% \\
		& SomnoNet & 88.0 & 100\% \\
		\bottomrule
	\end{tabular}
	\label{tab:nano_acc}
\end{table}

Table~\ref{tab:nano_size} compares parameter sizes with recent models. SomnoNet-Nano uses only 0.049M total parameters for inference and trains only about 0.013M parameters in the compact temporal head. It is therefore smaller than recent compact student-style models such as Distill-S, highlighting its potential for deployment-oriented applications where both storage footprint and adaptation cost are important.

\begin{table}[!t]
	\centering
	\caption{Model-size comparison between SomnoNet variants and representative public baselines. For SomnoNet-Nano, total parameters describe inference size, whereas trainable parameters describe compact-head training cost.}
	\setlength{\tabcolsep}{3pt}
	\begin{tabular}{ccc}
		\toprule
		Method & Total Params & Trainable Params \\
		\midrule
		SomnoNet-Nano (ours) & \textbf{0.049M} & \textbf{0.013M} \\
		Distill-S\cite{park2025distillsleep} & \underline{0.109M} & \underline{0.109M} \\
		SomnoNet (ours) & 0.43M & 0.43M \\
		SalientSleepNet\cite{SalientSleepNet} & 0.9M & 0.9M \\
		U-Time\cite{U-time} & 1.1M & 1.1M \\
		TinySleepNet\cite{TinySleepNet} & 1.3M & 1.3M \\
		SleepEEGNet\cite{SleepEEGNet} & 2.1M & 2.1M \\
		XSleepNet\cite{XSleepNet} & 5.6M & 5.6M \\
		Distill-T\cite{park2025distillsleep} & 6.45M & 6.45M \\
		DeepSleepNet\cite{DeepSleepNet} & 21M & 21M \\
		\bottomrule
	\end{tabular}
	\label{tab:nano_size}
\end{table}

To provide a more practical assessment of computational efficiency, Table~\ref{tab:efficiency} reports parameter count, FLOPs, and inference latency for different temporal modules using the same encoder on the first fold of Physio2018. The Bi-GRU variant achieved the best accuracy among the tested temporal modules, whereas SomnoNet-Nano delivered the best overall efficiency. To make the timing results reproducible, both GPU and CPU latency were measured with batch size 1 under FP32 precision.

\begin{table}[!t]
	\centering
	\caption{Efficiency and temporal-module comparison on the first fold of Physio2018. GPU latency was measured on an NVIDIA RTX 4090, and CPU latency was measured on an Intel Core i7-12700F. All latency values are per 30-second epoch, with batch size 1 and FP32 precision. TF denotes the Transformer-based temporal module.}
	\footnotesize
	\setlength{\tabcolsep}{2pt} 
	\begin{tabular}{lccccc}
		\toprule
		Method & Params & FLOPs & GPU & CPU & Acc. \\
		& (k) & (M) & (ms) & (ms) & (\%) \\
		\midrule
		SomnoNet & 434.34 & 5802.42 & 3.81 & 232.55 & 81.5 \\
		SomnoNet-GRU & 260.77 & 5780.40 & 2.77 & 180.36 & 80.5 \\
		SomnoNet-TF & 759.36 & 6030.48 & 1.73 & 129.73 & 77.9 \\
		SomnoNet-Nano & \textbf{49.24} & \textbf{1039.59} & \textbf{1.09} & \textbf{29.49} & 80.5 \\
		\bottomrule
	\end{tabular}
	\label{tab:efficiency}
\end{table}

The core mechanism behind SomnoNet-Nano is the strategy of freezing the convolutional encoder while replacing the original multi-layer recurrent stack with a compact recurrent unit. This yields stable feature reuse, effective compression, and efficient training. More importantly, the comparison with SomnoNet-Nano-JT suggests that the proposed architecture is not merely compressible, but structurally efficient: the encoder appears to capture transferable local morphology particularly well, whereas directly shrinking and jointly training the temporal stack is less effective. The CPU benchmark further strengthens this deployment-oriented interpretation: SomnoNet-Nano reduces CPU latency from 232.55 ms to 29.49 ms compared with the full SomnoNet, corresponding to an approximately 7.9$\times$ speedup while retaining strong accuracy. Because the current architecture uses bidirectional context aggregation, the reported latency should be interpreted as an offline or buffered-inference benchmark rather than as a strict causal real-time acquisition-and-processing validation.

\section{Interpretability Analysis}

Although deep neural networks achieve high accuracy in medical sleep staging tasks, their ``black-box'' nature limits clinical trust. To improve interpretability, we analyze SomnoNet from three complementary levels: a \emph{voting-based decision model}, a \emph{feature vector-based decision model}, and a \emph{time series-based decision model}. These approaches are derived from the hierarchical structure of SomnoNet and are inspired by attribution-style visualization methods such as CAM and Grad-CAM\cite{zhou2016cam,Grad-cam}, but are adapted to 1D EEG temporal analysis rather than 2D spatial localization in image recognition. Importantly, these three levels are not intended as separate deployment systems for all users; instead, they provide complementary views of model evidence at different levels of abstraction, from simple rhythm-level summaries to sequential temporal attribution. At the same time, we do not claim model-agnostic generality here, since the present study did not evaluate the same interpretability pipeline on alternative backbone architectures.

\subsection{Fixed Simple Decision Head (Voting-based)}

\begin{figure}[!t]
	\centering
	\includegraphics[width=3.2in]{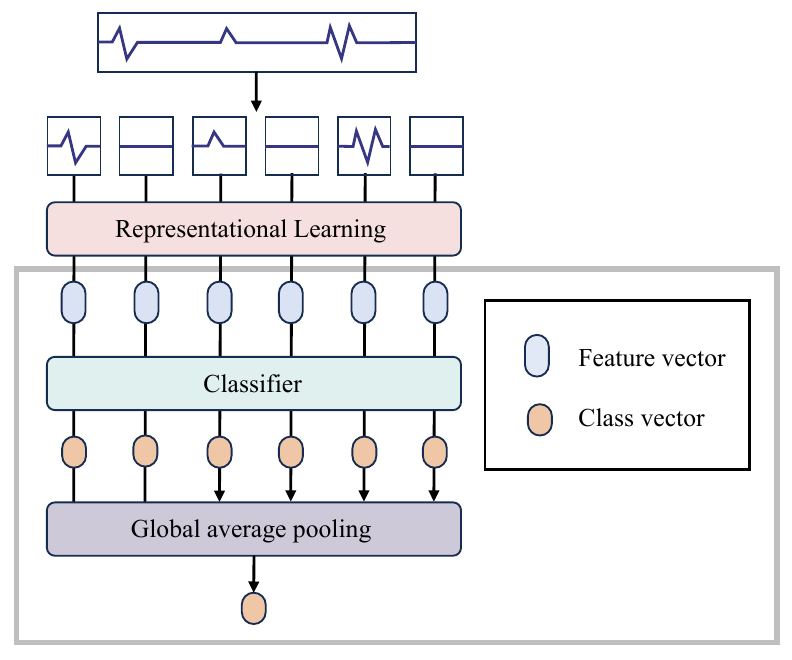}
	\caption{Voting-based interpretability structure. The classifier independently predicts sleep stages for each of the $N$ feature vectors extracted from EEG sub-segments within a 30-second sleep frame. The final sleep stage label for the frame is obtained via global average pooling, effectively implementing a ``voting'' mechanism across sub-segments.}
	\label{fig:inter_1}
\end{figure}

As illustrated in Fig.~\ref{fig:inter_1}, at the first level we adopt a macro-temporal perspective by dividing a sleep segment $X^t$ into $N$ sub-segments $x_i^t$. Each sub-segment is independently processed through feature extraction and classification, and the final prediction of the entire sleep frame is obtained via global average pooling (voting), as expressed in Eq.~\ref{eq:infer1}:

\begin{equation}
	\begin{aligned}
		\left\{ x_{i}^{t} \right\}_{i=1}^{N} & = \operatorname{Chunk}\left( X^{t} \right),                 \\
		f_{\text{rep},i}^{t}                 & = \operatorname{Rep}\left( x_{i}^{t} \right),               \\
		c_{i}^{t}                            & = \operatorname{Classifier}\left( f_{\text{rep},i}^{t} \right), \\
		c^{t}                                & = \frac{1}{N}\sum_{i=1}^{N} c_{i}^{t},                      \\
		y^{t}                                & = \operatorname{Argmax} \left( c^{t} \right).              
	\end{aligned}
	\label{eq:infer1}
\end{equation}

Here, $c_i^t$ denotes the classification result of the $i$-th sub-segment, and $c^t$ represents the predicted class distribution of the entire EEG frame. This design converts frame-level prediction into a voting mechanism among sub-segments, where each segment contributes one vote and the final prediction reflects the combined influence of different temporal portions. By analyzing the distribution of $\{c_i^t\}_{i=1}^{N}$, one can intuitively identify the critical time segments that drive the prediction, providing a coarse-grained interpretability pathway.

\begin{figure}[!t]
	\centering
	\includegraphics[width=3.2in]{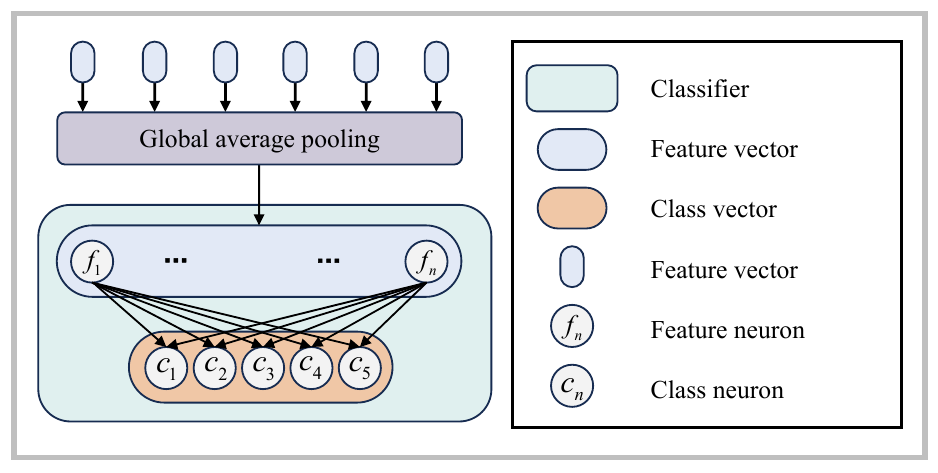}
	\caption{Feature vector-based interpretability structure. Feature vectors extracted from the $N$ EEG sub-segments within a 30-second sleep frame are first aggregated via global average pooling to obtain a single frame-level feature vector representing the entire sleep frame. This vector is then passed through the classifier (fully connected layer) to produce the predicted sleep stage.}
	\label{fig:inter_2}
\end{figure}

\subsection{Extended Decision Head (Feature Vector-based)} 

As illustrated in Fig.~\ref{fig:inter_2}, building upon the voting-based decision model, we propose a feature vector-based decision analysis framework, which reveals the decision logic of the model by analyzing its sensitivity to feature vectors. This framework supports forward propagation through a single-layer fully connected classifier and normalized backward propagation, and can be extended to more complex classifier structures such as multi-layer fully connected networks or recurrent networks.

\subsubsection{Forward Propagation for Decision Vectors}

Each segment feature vector $f_{\text{rep},i}^t$ is globally aggregated to form a frame-level feature vector $f_{\text{global}}^t$, which is then passed through a fully connected classifier to produce the prediction, as shown in Eq.~\ref{eq:infer2}:  

\begin{equation}
	\begin{aligned}
		f_{\text{global}}^{t} &= \frac{1}{N}\sum_{i=1}^{N} f_{\text{rep},i}^{t},\\
		c^{t} &= \operatorname{Classifier}(f_{\text{global}}^{t}),\\
		y^{t} &= \operatorname{Argmax}\left(c^{t}\right).
	\end{aligned}
	\label{eq:infer2}
\end{equation}

The computation of a single-layer fully connected classifier is expressed in Eq.~\ref{eq:infer2_2}, where $\mathbf{W}_{\text{fc}}$ and $\mathbf{b}_{\text{fc}}$ denote the weight matrix and bias vector of the fully connected layer, respectively. Let $L$ be the length of $f_{\text{global}}^{t}$ and $C$ the length of $c^{t}$; then $\mathbf{W}_{\text{fc}}$ is a $C \times L$ matrix and $\mathbf{b}_{\text{fc}}$ is a $C$-dimensional vector.

\begin{equation}
	\begin{aligned}
		\operatorname{Classifier}\left(f_{\text{global}}^{t}\right) = f_{\text{global}}^{t}\mathbf{W}_{\text{fc}}^{\top} + \mathbf{b}_{\text{fc}}.
	\end{aligned}
	\label{eq:infer2_2}
\end{equation}

The forward propagation decision vector is computed as in Eq.~\ref{eq:infer2_3}. Here, $Att_{f\_\text{global}}^{t}$ represents the contribution vector of the global frame feature $f_{\text{global}}^{t}$ to the predicted class $\text{pred}^t$, of length $L$. The vector $\mathbf{W}_{\text{fc}}[\text{pred}^t]$ is the row of the weight matrix corresponding to $\text{pred}^t$, and $\mathbf{b}_{\text{fc}}[\text{pred}^t]$ is the corresponding bias term (fixed during inference). Additionally, $Att_{f}^{t}$ denotes the decision vectors of the feature set $\{f_i^t \mid i=1,\dots,N\}$, and $Att_{x}^{t}$ denotes the decision vectors of the sample sub-segments $\{x_i^t \mid i=1,\dots,N\}$. For simplicity, the bias term $\mathbf{b}_{\text{fc}}[\text{pred}^t]$ is omitted in the subsequent approximation.

\begin{equation}
	\begin{aligned}
		Att_{f\_\text{global}}^{t} &= f_{\text{global}}^{t}\cdot \mathbf{W}_{\text{fc}}[\text{pred}^t]^{\top} + \mathbf{b}_{\text{fc}}[\text{pred}^t] \\
		&\sim f_{\text{global}}^{t}\cdot \mathbf{W}_{\text{fc}}[\text{pred}^t]^{\top},\\[4pt]
		Att_{x}^{t} = Att_{f}^{t} &= \frac{1}{N}\sum_{i=1}^{N}\left\{ \frac{f_{i}^{t}}{f_{\text{global}}^{t}} \cdot Att_{f\_\text{global}}^{t} \right\} \\
		&\sim \frac{1}{N}\sum_{i=1}^{N}\left\{ f_{i}^{t}\cdot \mathbf{W}_{\text{fc}}[\text{pred}^t]^{\top} \right\}.
	\end{aligned}
	\label{eq:infer2_3} 
\end{equation}

Using this formulation, we obtain forward interpretability results for the single-layer fully connected classifier.

\subsubsection{Backward Propagation for Decision Vectors}

For more complex classifiers, such as multi-layer fully connected networks or recurrent neural networks, relying solely on forward propagation may not fully reveal the true contribution of input features to the prediction. To address this, we adopt a backward propagation approach to quantify the sensitivity of each input feature to the final prediction. The forward process of the model can be formalized as in Eq.~\ref{eq:infer2_4}:

\begin{equation}
	\begin{aligned}
		f_{\text{rep},i}^{t} &= \operatorname{Rep}(x_{i}^{t}), \\ 
		c^{t} &= \operatorname{Classifier}\big(\{ f_{\text{rep},i}^{t} \mid i=1,\ldots,N \}\big), \\ 
		y^{t} &= \operatorname{Argmax}(c^{t}). 
	\end{aligned}
	\label{eq:infer2_4}
\end{equation}

Each sample block $x_{i}^{t}$ is first processed by the feature extractor $\operatorname{Rep}$ to obtain the corresponding feature vector $f_{\text{rep},i}^{t}$. The set of feature vectors $\{ f_{\text{rep},i}^{t} \mid i = 1,\ldots,N \}$ within a single sleep frame is then passed through the classifier $\operatorname{Classifier}$ to yield the frame-level classification vector $c^{t}$, and the final prediction $y^{t}$. Based on this formulation, the decision vectors can be further derived as in Eq.~\ref{eq:infer2_5}.

\begin{equation}
	\begin{aligned}
		g^{t} &= \frac{\partial c^{t}[\text{pred}]}{\partial f_{\text{set}}^{t}},      \\
		a_{i}^{t} &= \frac{1}{L}\sum_{l=1}^{L} f_{\text{set},i,l}^{t} g_{i,l}^{t},\quad i=1,\ldots,N,\\
		Att_{x}^{t} &= \operatorname{Norm}\left(\{a_{i}^{t}\}_{i=1}^{N}\right),
	\end{aligned}
	\label{eq:infer2_5}
\end{equation}

where $f_{\text{set}}^{t} = \{ f_{\text{rep},i}^{t} \mid i = 1,\ldots,N \}$ denotes the collection of feature vectors within a sleep frame. If $L$ is the dimensionality of $f_{\text{rep},i}^{t}$, then $f_{\text{set}}^{t}$ can be viewed as an $N \times L$ feature matrix. The gradient $g^{t}$ measures the sensitivity of the predicted-class logit to each feature dimension. The segment-level score $a_i^t$ is computed by gradient--activation weighting and then normalized by $\operatorname{Norm}(\cdot)$ for visualization. This produces an $N$-dimensional decision vector $Att_x^t$ that quantifies the relative contribution of each sample block to the final prediction without introducing an arbitrary sign reversal.

\subsection{Sequence-aware Decision Head (Time Series-based)}  

\begin{figure}[!t]
	\centering
	\includegraphics[width=3.2in]{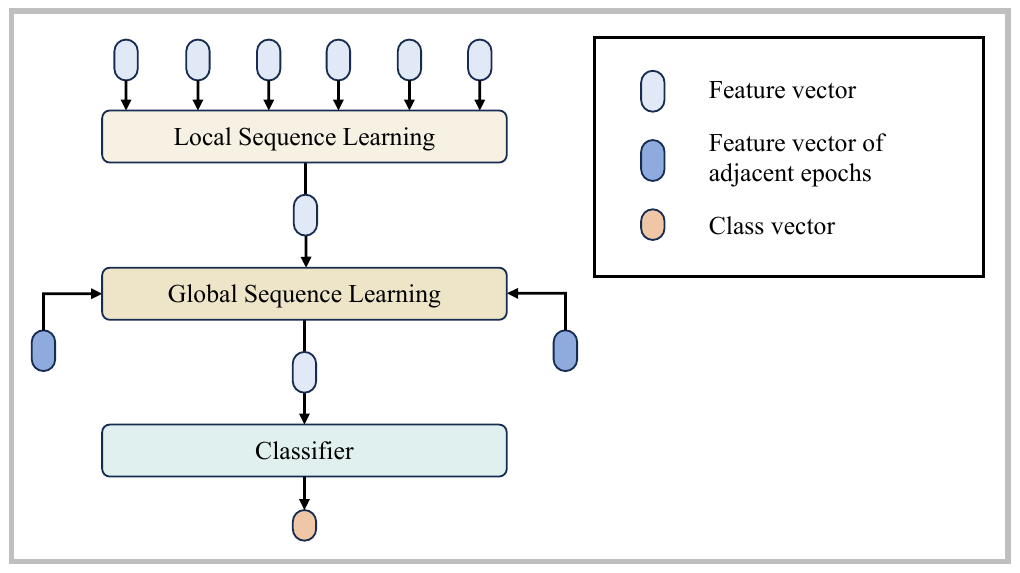}
	\caption{Time series-based interpretability structure. Feature vectors extracted from the $N$ EEG sub-segments within a 30-second sleep frame are first processed by the local sequence learning module to capture intra-frame temporal dependencies. The resulting frame-level representations are then fed into the global sequence learning module, which models inter-frame dependencies across adjacent sleep frames. Finally, the classifier outputs the predicted sleep stage for the current frame, enabling a comprehensive visualization of how both intra-frame and inter-frame temporal dynamics contribute to the model's decision-making process.}
	\label{fig:inter_3}
\end{figure}

Sleep staging depends not only on single-frame features but also on temporal dependencies across consecutive frames. While the voting-based and feature vector-based attribution methods reveal within-frame importance, they cannot capture inter-frame relationships. To address this, we introduce temporal dynamic modeling (Fig.~\ref{fig:inter_3}), keeping the feature extractor frozen while training a lightweight recurrent neural network (RNN) to capture sequential evolution.

During forward propagation, the local sequence learning module $\operatorname{Seq}_{\text{local}}$ first integrates all sample block features $\{ f_{\text{rep},i}^{t} \}_{i=1}^N$ within a single sleep frame $X^t$, yielding a temporally enriched global feature vector $f_{\text{global}}^{t}$. This is further combined with the forward feature of the previous frame $\overrightarrow{f_{\text{seq}}^{(t-1)}}$ and the backward feature of the next frame $\overleftarrow{f_{\text{seq}}^{(t+1)}}$ via a global sequence learning module to obtain comprehensive sequential features:

\begin{equation}
	\begin{aligned}
		f_{\text{global}}^{t} &= \operatorname{Seq\_local}\big( \{ f_{\text{rep},i}^{t} \mid i=1,\ldots,N \} \big), \\
		f_{\text{seq}}^{t} &= \operatorname{Seq\_global}\big(\overrightarrow{f_{\text{seq}}^{(t-1)}}, f_{\text{global}}^{t}, \overleftarrow{f_{\text{seq}}^{(t+1)}}\big), \\
		c^{t} &= \operatorname{Classifier}(f_{\text{seq}}^{t}), \\
		y^{t} &= \operatorname{Argmax}(c^{t}).
	\end{aligned}
	\label{eq:infer_3}
\end{equation}

By extending Eq.~\ref{eq:infer2_5}, we derive the sequence-aware decision vector $Att_{x}^{t}$ using the same gradient-based approach. In contrast to the previous two methods, this approach explicitly incorporates temporal dependencies within the sequence, enabling more accurate inference of sleep states. By modeling the temporal order in the data, the time series-based model can better capture dynamic patterns and evolving trends, thereby improving adaptability to non-stationary EEG signals.

Using the first fold of cross-validation on the Physio2018 dataset, Table~\ref{tab:inter_acc} compares sleep staging accuracy across the three interpretability approaches. The results indicate that the voting-based and feature vector-based models achieve relatively lower overall accuracy, as they do not fully leverage local and global temporal information. Despite structural differences, both models have limited utilization of input features during forward propagation, resulting in comparable accuracy.

\begin{table}[!t]
	\centering
	\caption{Accuracy comparison of different interpretability models on the first Physio2018 fold.}
	\begin{tabular}{cc} 
		\toprule
		Method      & Accuracy (\%) \\ \midrule
		Voting-based model & 73.2     \\
		Feature vector-based model & 73.2     \\
		Time series-based model & 81.5     \\ \bottomrule
	\end{tabular}
	\label{tab:inter_acc}
\end{table}

To quantitatively examine whether the highlighted temporal segments are functionally related to model decisions, we further performed a segment-feature deletion test on the same first Physio2018 fold used for the sampling-rate and interpretability analyses. For each correctly classified test epoch, segment attribution was computed with respect to the originally predicted class. We then progressively replaced the top-ranked segment features with a baseline feature vector and measured the decrease in the original-class confidence. Random deletion, averaged over 20 random masks, and bottom-attribution deletion were used as controls. The deletion ratios were set to 10\%, 20\%, 30\%, 40\%, and 50\%, and the area over the perturbation curve (AOPC) was used as the main faithfulness metric.

\begin{table}[!t]
	\centering
	\caption{Attribution faithfulness evaluated by segment-feature deletion on correctly classified epochs from the first Physio2018 fold. AOPC and $\Delta$Conf@20\% denote the average original-class confidence drop across deletion ratios and at 20\% deletion, respectively. Flip@20\% reports the percentage of samples whose predicted class changed after 20\% deletion.}
	\footnotesize
	\setlength{\tabcolsep}{3pt}
	\begin{tabular}{lcccc}
		\toprule
		Deletion & AOPC$\uparrow$ & Macro$\uparrow$ & $\Delta$Conf$\uparrow$ & Flip$\uparrow$ \\
		 &  & AOPC & @20\% & @20\% \\
		\midrule
		Top & \textbf{0.1049} & \textbf{0.0965} & \textbf{0.0598} & \textbf{5.99} \\
		Random & 0.0496 & 0.0440 & 0.0274 & 2.26 \\
		Bottom & 0.0387 & 0.0367 & 0.0180 & 1.30 \\
		\bottomrule
	\end{tabular}
	\label{tab:deletion}
\end{table}

\begin{figure*}[!t]
	\centering
	\subfloat[]{\includegraphics[width=0.45\textwidth]{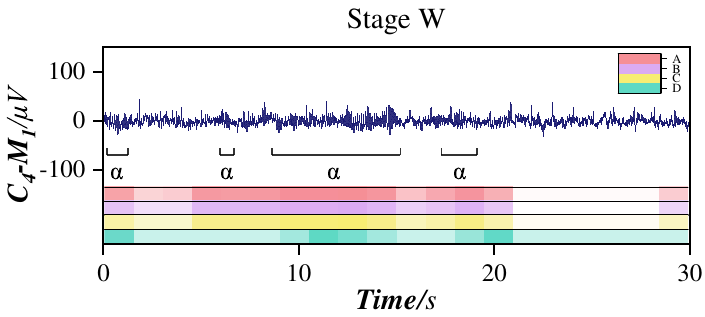}
		\label{fig:cam_6}}
	\hspace{-0.5em}
	\subfloat[]{\includegraphics[width=0.45\textwidth]{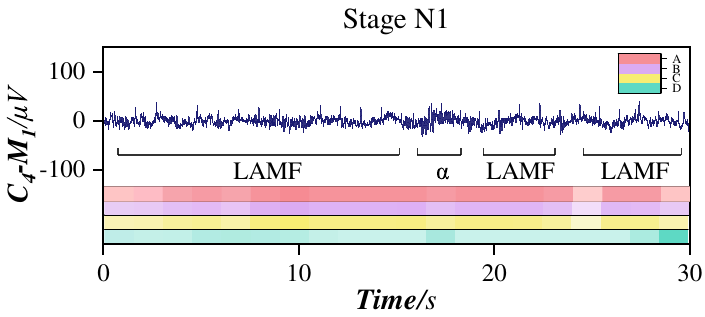}
		\label{fig:cam_8}}

	\subfloat[]{\includegraphics[width=0.45\textwidth]{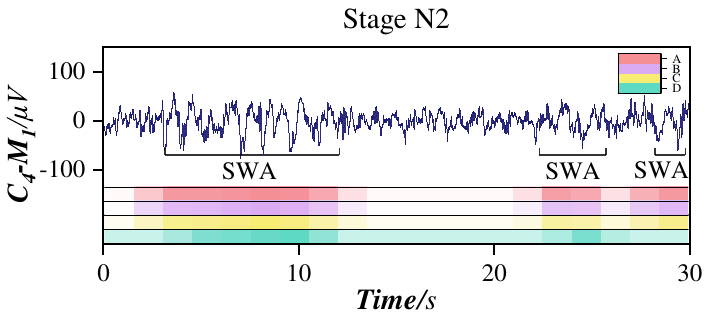}
		\label{fig:cam_70}}
	\hspace{-0.5em}
	\subfloat[]{\includegraphics[width=0.45\textwidth]{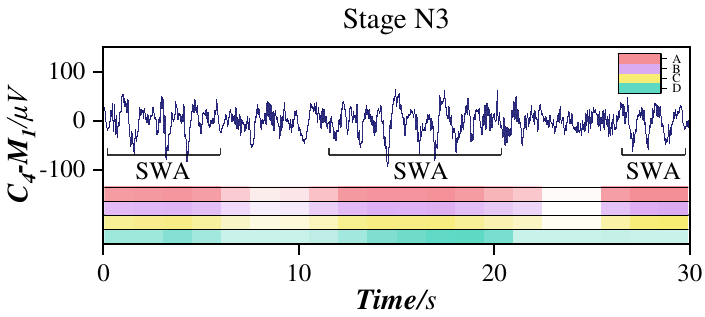}
		\label{fig:cam_76}}
	\caption{Interpretability analysis of automatic sleep staging using single-channel EEG ($C_3$--$A_2$) from the Physio2018 dataset. Each panel corresponds to a 30-second epoch: (a) Stage W, (b) Stage N1, (c) Stage N2, and (d) Stage N3. For each epoch, four interpretability approaches are illustrated: (A) voting-based decision model with a fixed simple decision head, (B) feature vector-based decision model (forward propagation), (C) feature vector-based decision model (backward propagation), and (D) time series-based decision model. The four heatmaps beneath each EEG trace indicate the decision attention derived from the corresponding methods, where darker colors denote higher model focus. Characteristic rhythms such as $\alpha$ activity and slow-wave activity (SWA) are annotated with brackets for clinical reference.}
	\label{fig:cam_all}
\end{figure*}

As shown in Table~\ref{tab:deletion}, deleting the top-attribution segments produced a much larger decrease in original-class confidence than random or bottom-attribution deletion. The top-deletion AOPC was approximately 2.1$\times$ that of random deletion and 2.7$\times$ that of bottom deletion. The same trend was observed for macro-class AOPC, indicating that the result was not dominated by majority stages. At 20\% deletion, top-attribution deletion reduced confidence by 0.0598 and changed 5.99\% of predictions, compared with 0.0274 and 2.26\% for random deletion. These results suggest that the rhythm-aware attribution maps are not merely visually plausible but are also decision-relevant for SomnoNet.

To further analyze model behavior, sleep frames 6, 8, 70, and 76 from subject tr03-0005 in Physio2018 (corresponding to Stages W, N1, N2, and N3) are visualized via decision vector heatmaps (Fig.~\ref{fig:cam_all}) for single-channel EEG (central $C_3$--$A_2$ lead). Heatmaps A, B, C, and D correspond to the voting-based model, the forward feature vector-based model, the backward feature vector-based model, and the time series-based model, respectively. Color intensity reflects the degree of attention on each feature, providing an intuitive visualization of the model's decision-making process.

Observations include:

\begin{itemize}
	\item \textbf{Stage W frame (Fig.~\ref{fig:cam_6}):} All four models primarily attend to $\alpha$ rhythms, consistent with AASM scoring guidelines. The lighter color in method (D) indicates greater emphasis on feature vectors from adjacent sleep frames.
	\item \textbf{Stage N1 frame (Fig.~\ref{fig:cam_8}):} The voting-based and feature vector-based models (A, B, C) focus on low-amplitude mixed-frequency (LAMF) rhythms. In contrast, the time series-based model (D) emphasizes $\alpha$ rhythms and does not fully capture the LAMF rhythm. The lighter color again suggests that method (D) incorporates information from neighboring frames.
	\item \textbf{Stage N2 and Stage N3 frames (Figs.~\ref{fig:cam_70} and \ref{fig:cam_76}):} All models correctly detect slow-wave activity. In method (D), the lighter color indicates a stronger contribution from adjacent frames.
\end{itemize}

Further comparison between decision vectors obtained from the forward method (B) and the backward propagation method (C) demonstrates a high degree of visual consistency, supporting the use of gradient-based attribution for this hierarchical EEG model. Nevertheless, this analysis should be interpreted as a structure-aware explanation for SomnoNet-like architectures rather than as a model-agnostic interpretability paradigm.

Overall, the sequence-aware model achieves the highest classification accuracy; however, its reliance on adjacent frames can reduce the visual prominence of distinctive local rhythms. In contrast, the non-sequential voting-based and feature vector-based heads, although less accurate, more directly highlight characteristic local patterns. This comparison suggests a practical trade-off between context-driven accuracy and local-rhythm transparency, which is important when designing clinically inspectable sleep staging systems.

\section{Conclusion}

In this study, we presented \textbf{SomnoNet}, a hierarchical end-to-end framework for automatic sleep staging from raw single-channel EEG. The model is motivated by clinical sleep scoring practice and integrates multi-scale local representation learning with intra-epoch and inter-epoch temporal modeling. On two large-scale public datasets, the full model achieved competitive performance while preserving a clinically meaningful architectural rationale.

We further designed \textbf{SomnoNet-Nano} as a deployment-oriented compact variant. By reusing the encoder learned by the full model and replacing the original hierarchical temporal stack with a lightweight sequence unit, SomnoNet-Nano substantially reduces parameter count and computational cost while retaining most of the full-model accuracy. Under FP32 inference, SomnoNet-Nano processes a 30-second epoch in 29.49 ms on an Intel Core i7-12700F CPU, supporting its practicality for buffered deployment scenarios. The frozen-encoder result also provides evidence for the structural efficiency of the proposed encoder: local EEG morphology learned by the full model remains useful even when the temporal head is aggressively compressed.

In addition, we introduced rhythm-aware decision analysis to relate model predictions to characteristic EEG rhythms and temporal contexts. This analysis provides complementary views of model evidence, from local rhythm-level summaries to sequence-aware attribution. The representative-fold deletion test further shows that the highlighted segments are functionally related to model confidence, thereby improving the transparency of SomnoNet-based sleep staging.

Several limitations should be noted. First, all experiments were retrospective and based on public datasets; prospective clinical validation remains necessary before clinical deployment. Second, SomnoNet uses bidirectional context and is therefore best suited to offline or buffered inference rather than strictly causal streaming. Third, SomnoNet-Nano requires a pretrained full encoder, which should be considered when evaluating training cost. Finally, although the proposed interpretation strategy links model evidence to rhythm-level EEG patterns and is supported by a representative-fold deletion test, its clinical utility should be further assessed with sleep experts and larger-scale fold-wise and cross-dataset faithfulness analyses.

Future work will investigate causal or low-latency variants for online use, broader multimodal extensions, larger-scale expert validation of attribution maps, and deployment validation in home-based and clinical sleep monitoring environments.

\section*{Code Availability}
The implementation is publicly available at \url{https://github.com/komdec/SomnoNet.git}.

\section*{Declarations}
\noindent\textit{Author Contributions}

S.G. designed the study, conducted all experiments, and wrote the manuscript. G.S. provided resources, supervised the project, and offered overall guidance throughout the study. All authors reviewed and approved the final manuscript.

\noindent\textit{Acknowledgment}

The authors would like to thank the creators and maintainers of the open-source datasets used in this study for making their resources publicly available.

\noindent\textit{Funding}

The authors received no specific funding for this work.

\noindent\textit{Conflict of Interest}

The authors declare that they have no conflict of interest.

\section*{References}
\bibliographystyle{IEEEtran}
\bibliography{sleep_bibliography}

\begin{IEEEbiography}
	[{\includegraphics[
		width=1.0in,
		height=1.2in,
		clip,
		keepaspectratio
		]{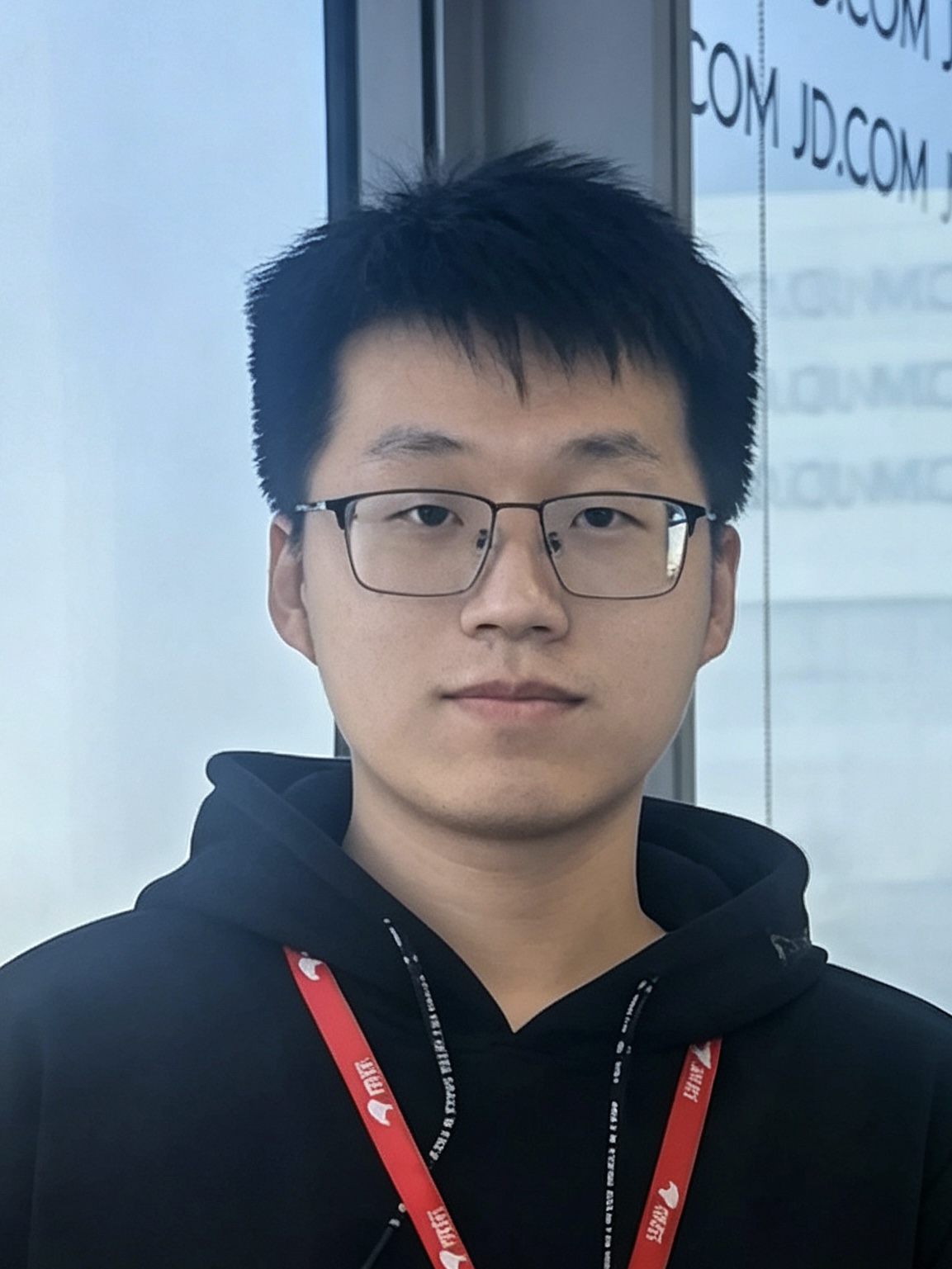}}]
	{Shengwei Guo}
	received the B.E. degree in automation and the M.Eng. degree in electronic information from Heilongjiang University, Harbin, China, in 2020 and 2024, respectively. He is currently with the Key Laboratory of Information Fusion Estimation and Detection, School of Electronic Engineering, Heilongjiang University. His research interests include automatic sleep staging, neurophysiological signal processing, computer vision, and robotics.
\end{IEEEbiography}

\vskip -2\baselineskip plus -1fil

\begin{IEEEbiography}
	[{\includegraphics[
		width=1.0in,
		height=1.2in,
		clip,
		keepaspectratio
		]{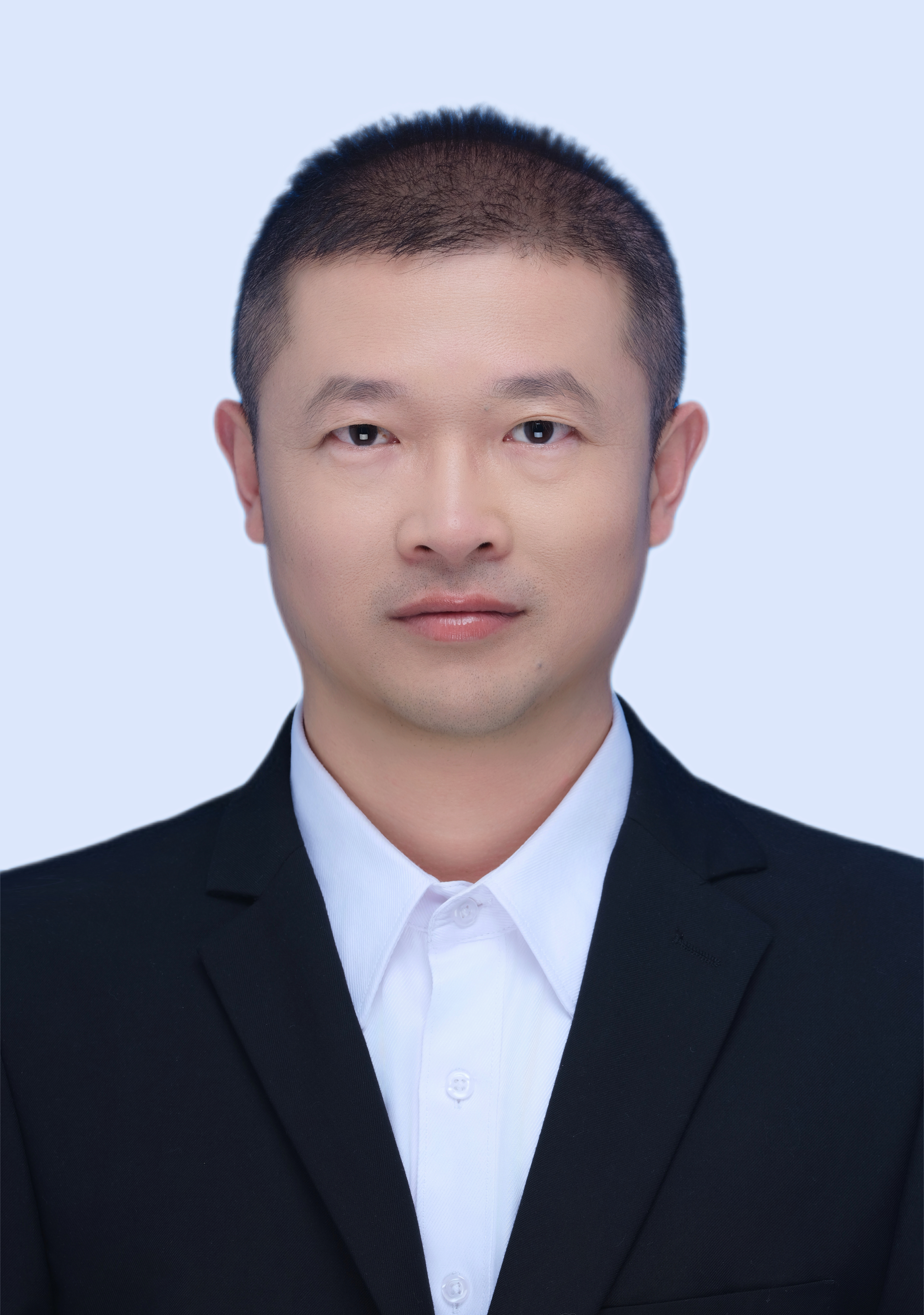}}]
	{Guobing Sun}
	received the B.E. and Ph.D. degrees from Harbin Institute of Technology, Harbin, China, in 2002 and 2009, respectively. He is currently an Associate Professor with the Key Laboratory of Information Fusion Estimation and Detection, School of Electronic Engineering, Heilongjiang University, Harbin, China. His research interests include biomedical signal processing, machine learning, computer vision, and intelligent control.
\end{IEEEbiography}

\end{document}